\documentclass[prd,a4paper,aps,epsfig,showpacs,twocolumn,floats]{revtex4}

\newcommand{\be}{\begin{equation}}
\newcommand{\ee}{\end{equation}}
\newcommand{\bea}{\begin{eqnarray}}
\newcommand{\eea}{\end{eqnarray}}
\newcommand{\mn}{{\mu\nu}}

\begin{document}
\title{DSR as an explanation of cosmological structure}
\author{Jo\~ao  Magueijo}
\affiliation{
Theoretical Physics Group, Imperial College, London, SW7 2BZ}
\date{\today}
\begin{abstract}
{Deformed special relativity (DSR) is one of the possible realizations of
a varying speed of light (VSL). It deforms the usual quadratic dispersion
relations so that the speed of light becomes energy dependent,
with preferred frames avoided by postulating a non-linear
representation of the Lorentz group. The theory may be used to induce
a varying speed of sound capable of generating
(near) scale-invariant density fluctuations, as discussed in a recent Letter.
We identify the non-linear representation of the Lorentz group that leads
to scale-invariance, finding a universal result. We also examine the higher
order field theory that could be set up to represent it.}
\end{abstract}
\pacs{0000000}
\maketitle

\bigskip

\section{Introduction}
In a recent Letter~\cite{csdot}  we proposed a mechanism for producing
scale-invariant density fluctuations of appropriate amplitude
based on a decaying speed of sound. The mechanism
is quite general and can be implemented using a variety of methods.
The examples of $\kappa$-essence~\cite{kappa,garriga}
and varying speed of light (VSL)~\cite{vsl,jm,am}
were given, and other possibilities,
such as non-adiabatic hydrodynamical matter, were considered.
But in~\cite{csdot}
we emphasized the generality of the proposal and deliberately chose
not to marry it to any specific model.
This attitude was reversed in~\cite{csbim}, where we initiated model-building
appealing to bimetric VSL theories.
Here we propose yet another concrete VSL realization of scale-invariant
fluctuations, this time
based on theories that rub shoulders with the phenomenology
of quantum gravity.

In spite of the strong reactions they elicit~\cite{caus,ellis,reply}
VSL theories may be nothing more offensive than non-trivial realizations of the
Lorentz group. Two such approaches stand out:
bimetric VSL  and deformed special relativity (DSR).
In the former the Lorentz group is realized by different metrics for
matter and for gravity, leading to different speeds for photons
(and other massless matter particles) and gravitons~\cite{bim}.
In the latter the dispersion relations are deformed from their usual
quadratic form, so that the  speed of all massless particles
becomes energy dependent~\cite{dsr}.
To prevent the introduction of a preferred frame one then chooses a suitable
non-linear representation of the Lorentz group~\cite{ljprl,ljprd}.
Other interpretations are possible---and of importance---but
they won't be relevant in this paper~\cite{desitter,2d,nc,nc1}.

It is not evident that DSR may be used to implement the varying
speed mechanism of~\cite{csdot}, where the propagation speed $c_s$
is envisaged  to vary with time, not with energy. In bimetric
theories~\cite{csbim} the speed of light varies in time and this
is passed on to all matter propagation speeds, including the speed
of acoustic oscillations, $c_s$. But in DSR the speed of light is
energy or wavelength dependent: not time-dependent. Yet DSR can
indeed be used to implement a time-varying $c_s$ ``by proxy'', in
an expanding Universe. If we focus on a fixed comoving mode (as
done in the usual perturbations' calculation) we see its physical
size stretch in time. Its ``energy scale'' therefore changes in
time, and so, under DSR, the mode is effectively subject to a time
varying speed of light and consequently of sound.

The cosmological redshift acts to convert
a frequency dependent speed of light into a time-dependent speed of light.
This idea was already recognized in~\cite{steph}, in a different guise.
If the speed of massless particles increases with energy
then the fact that the ``average particle'' has a higher energy
in the early Universe means that the ``ambient'' speed of light
is also higher.

\section{Very basic DSR}
We start with a quick review of DSR, cast in a formalism that can
be used here. It's possible that alternative
formulations~\cite{nc,nc1} may be plugged into the fluctuations'
calculation that follows, but this has not been checked.

Let us consider a  deformed dispersion relation (DDR) of the form:
\be\label{ddr} E^2f_1^2-k^2 f_2^2=m^2 \ee where $f_1$ and $f_2$
may be general functions of $E$ and $k$. Then the group speed of
light (or any other massless particles) $c=\frac{dE}{dp}$ becomes
energy dependent. Expression (\ref{ddr}) is not invariant under
the linear Lorentz transformations. However it is possible to find
a non-linear representation of the Lorentz group which leaves it
invariant and prevents the introduction of a preferred frame, so
that relativity between inertial observers is preserved. This is
done by considering the map: \be U \circ (E, {\bf p})=(Ef_1,{\bf
p}f_2), \label{udef} \ee and then changing the representation
according to:
\begin{equation}\label{U}
K^i = U^{-1} L_0^{\ i} U
\end{equation}
where $L_{ab} = p_a {\partial \over \partial p^b} -
 p_b {\partial \over \partial p^a}$ are the standard Lorentz generators.
Exponentiation then gives us a set of non-linear transformations
for which (\ref{ddr}) is invariant.

DSR has provided an excellent bridge between phenomenology and
quantum gravity. The existence of an invariant length or energy
scale is the central connection, and this can easily be
represented by the singular points of the transformation $U$. For
this reason such theories are also called ``doubly special''
relativity (fortunately leading to the same acronym).  However one
may look at it more generally. For a recent review the reader can
consult~\cite{giovrev}.

A number of issues arise at once. The theory is defined in
momentum space, and once linearity is lost duals no longer
decouple and mimic each other. The introduction of an energy
dependent space-time metric may then be necessary, the so-called
rainbow metric~\cite{rainbow}. But other constructions are
possible~\cite{leerain,sabreal,realdag,mig}. Also the field theory
realization of DSR remains an open issue~\cite{nc,nc1}. Higher
order derivative (HOD) theories will be advocated here as a
realization of DSR~\cite{hod}, not so much because we love them,
but because they permit a direct realization of the varying speed
of sound mechanism, which we stress is the central topic in this
paper.

HOD theories have been much maligned, not always fairly. They are
usually blamed for hiding ghosts, but the argument is seldom
understood, often not even known to those who wield it. Many
pathologies used to smear HOD theories only occur if a theory with
infinitely many derivatives is truncated~\cite{barn,sabhod}. Also
the threat of pathologies doesn't even arise if the HOD behaviour
is reserved to spatial derivatives: then all that happens is that
more data is required at spatial infinity for the problem to be
well posed. This is definitely the case for the HOD theories
considered in this paper.

Note that various DSRs and HOD theories may be made to correspond
to a given DDR, particularly in the massless limit. The expressions
for $f_1$ and $f_2$ in (\ref{ddr}) could be changed (and they could be
seen as different functions of $E$ and $k$) leading to an algebraically
equivalent DDR but a different $U$ map, and so a different DSR and
HOD field theory.

\section{An adaptation of the decaying speed of sound mechanism}
It is straightforward to adapt the calculation in~\cite{csdot} to
the present, slightly different context (this was already done,
with a different motivation,  by~\cite{amend,piao}). Here we
predict ``from first principles'', by means of DSR, deformed
dispersion relations for the fluctuations. Thus, their speed of
sound is wavelength dependent. But since the wavelength of each
mode increases due to cosmic expansion, its speed of sound is time
dependent. This time dependence comes about by proxy---due to the
indirect effect of expansion---but it has equivalent effects in
terms of structure formation.

Let us consider DDRs in the form (\ref{ddr}) with $f_1=1$ and
$f_2=g(\lambda k)$, where $k$ is a wavevector and $\lambda$ a
parameter with units of $M^{-1}$. We can then take the massless
limit $m=0$ and adapt to an expanding universe by introducing
comoving $\omega$ and $k$, with $g\rightarrow g(\lambda k/a)$.
Thus: \be \omega=kg(\lambda k/a) \ee and if we consider an
asymptotic regime where $g$ behaves as the power-law $g(x)\propto
x^\gamma$ we find: \be
c=\frac{d\omega}{dk}=(\gamma+1)\frac{\omega}{k}\propto
{\left(\frac{\lambda k}{a}\right)}^\gamma\; . \ee Since
$a\propto\eta^\frac{1}{\epsilon-1}$ with
$\epsilon=\frac{3}{2}(1+w)$, where $w=p/\rho$ is the equation of
state, we can make contact with the law $c\propto \eta^{-\alpha}$
used in~\cite{csdot}, with \be \alpha=\frac{\gamma} {\epsilon
-1}\; .\ee But note that $c$ is now $k$ dependent too, an
important difference with respect to~\cite{csdot}.

Whether we employ a hydrodynamical fluid or a scalar field the
density fluctuations are described by a modified harmonic
oscillator equation. In terms of the curvature perturbation
$\zeta=-v/z$ its equation takes the
form~\cite{mukh,lidsey,garriga}: \be\label{veq} v''+\left[\omega^2
-\frac{z''}{z}\right]v=0 \ee but now $z \propto a$, as explained
in~\cite{amend} (cf.~\cite{csdot}). Here we can write $\omega=c k$
for convenience, because for the models considered $\omega/k$ and
$d\omega/dk$ (the phase and group speeds) are the same up to a
factor of order one.

For modes to start oscillating and then freeze out (i.e. for the
horizon problem to be solved), we need the term in $\omega$ in
Eq.(\ref{veq}) to dominate first. For an expanding Universe with
$w>-1/3$ (for which $\eta$ is positive and increases from zero)
this requires $\alpha>1$, that is: \be\label{hor}
\gamma>\gamma_0=\epsilon -1=\frac{1+3w}{2}\; . \ee This should be
seen as the condition zero for our calculation to make sense.

As in~\cite{csdot} Eqn.~(\ref{veq}) can be transformed into a
Bessel equation, with a boundary condition obtained in the WKB
limit. Under (\ref{hor}) modes start inside the horizon (set by $c
k\eta\sim 1$), so that we can ignore the term in $z''/z$ and find
the the appropriately normalized WKB solution \be\label{bc1}
v\sim\frac{e^{ik\int c d\eta}}{\sqrt{c k}}\sim \frac{e^{-i \beta c
k \eta}}{\sqrt{c k}} \ee where $\beta=1/(\alpha-1)>0$. The full
solution to  Eqn.~\ref{veq} then becomes \be
v=\sqrt{\beta\eta}(AJ_\nu(\beta c k\eta)+BJ_{-\nu}(\beta c
k\eta))\; . \ee where $A$ and $B$ are $k$-independent numbers of
order 1. The order $\nu$ is given by~\footnote{This expression
cannot be simply read off from~\cite{csdot} because here $z\propto
a$, not $z\propto a/c $.} \be
\nu=\frac{3-\epsilon}{2(\gamma-\epsilon +1)} \ee if $-1/3<w<1$
(minus this expression if $w>1$).

The spectrum left outside the horizon can now be found. Since $c
\eta$ is a decreasing function of time, the negative order
solution is the growing mode, so that asymptotically we have:
\be\label{vout} v\sim \frac{\sqrt{\beta \eta}}{(c k \eta)^{\nu}}\;
. \ee A further adaptation of~\cite{csdot} is required because $c$
in this expression is $k$ dependent. Since $\zeta=-v/z$, and the
spectral index is defined from $k^3\zeta^2= A^2 k^{n_S-1}$, we
have: \be n_S-1=\frac{\epsilon(\gamma-2)}{\gamma-\epsilon +1}\; .
\ee Scale-invariance therefore requires $\gamma=2$ for all
equations of state which satisfy $-1/3<w<1$~\footnote{For $w=1$ we
have $n_S=4$ for all $\gamma$; for $w>1$ the answer is more
complex}. This also complies with (\ref{hor}) for all $w$ in this
range. We can make the spectrum as red as we want (with
$\gamma_0<\gamma<2$), but the bluest it gets is $n_S=1+\epsilon$,
for $\gamma\rightarrow \infty$.

As in~\cite{csdot} the amplitude of the fluctuations may be found
by refining the DDRs to $g=1+(\lambda k)^2$, enforcing the correct
low energy limit. The scale $\lambda$ is then responsible for the
amplitude $A$ in $k^3\zeta^2= A^2 k^{n_S-1}$. Straightforward
algebra leads to $A\sim 1/(\lambda M_{Pl})$, implying that
$\lambda\sim 10^5 L_{Pl}$.

It is somewhat surprising that scale-invariance is associated with
a universal law, with details such as the background equation of
state $w$ falling out of the result (see~\cite{csdot,csbim} for
similar miracles). However here this is only true if the
sub-horizon normalization is chosen to match a vacuum quantum
state. Should it be a thermal state, as discussed in~\cite{csdot},
we have for the spectral index: \be n_S-1=\frac{(\epsilon
-1)(\gamma-1)-2}{\gamma-\epsilon+1} \ee so that scale-invariance
now requires \be
\gamma=\frac{\epsilon+1}{\epsilon-1}=\frac{5+3w}{1+3w} \ee and the
condition for scale-invariance becomes $w$ dependent. In the range
under study ($-1/3<w<1$) it requires $\gamma>2$, for example for a
radiation background it requires $\gamma=3$.

\section{The associated DSR}
What can we learn about DSR from this calculation?
Foremost we have constrained the dispersion relations of the
fluctuations. Specifically, if vacuum quantum fluctuations are
responsible for the structure of the Universe, then for all $w$
the DDRs should be of the form: \be \label{theddr}\omega^2-k^2(1+(\lambda
k)^2)^2=m^2\ee with $\lambda\sim 10^5 L_{Pl}$. This is not a
truncation: quite the opposite. We're probing the $\lambda k\gg 1$
regime, so the term in $(\lambda k)^4$ should be the highest power
in the dispersion relation. Different lower powers are admissible,
for example the DDRs could equally be $\omega^2-k^2(1+\lambda
k)^4=m^2$. The speed of light profile could be
$c=1+(\lambda k)^2$ or $c=(1+\lambda
k)^2$ with the same effects for structure formation; but
terms in $(\lambda k)^4$ or $(\lambda k)^6$ in $c(k)$ are
excluded. However if thermal (as opposed to quantum vacuum)
fluctuations are behind the structure of the Universe
more general DDRs become possible.

These DDRs may be be incorporated into a variety of DSRs or other
similar such constructions and the assumptions of the calculation are
not totally insensitive to the specific realization. Since
structure formation only probes the massless limit, only the ratio
of $f_1$ and $f_2$ is constrained. $f_1$ and $f_2$ can then
be seen as functions of $k$, $E$ or both. These algebraic
rearrangements do not affect the DDRs themselves, but do affect
the DSR that contains them~\cite{ljprd}. They're also  reflected
in the associated field theory~\cite{hod,sabhod}.

Crucially in our calculation, the field theory should be such that
the deformations only affect spatial gradients, i.e. derivatives
$D_\mu=(g_\mn -n_\mu n_\nu)\nabla^\nu$ where to zeroth order
$n_\mu$ points along the cosmological time. Then the background
evolution in a perturbed expanding Universe isn't affected by the
deformation; also $z\propto a$ in the perturbation calculations
(see~\cite{amend} for details).

For example we could take a DSR generated by \be U \circ
(E,p)=(E,p\sqrt{(1+(\lambda p)^4)}\; .\ee Its associated HOD field
theory satisfies the Klein-Gordon equation: \be
[\partial_0^2-\partial_i^2(1+(\lambda\partial_i)^4)+m^2]\phi=0 \ee
A coupling to gravity could be chosen such that this became: \be
[\nabla_\mu\nabla^\mu +\lambda^4 (D_\mu D^\mu)^2 +m^2]\phi=0\ee
(see~\cite{rainbow,finsler} for possible couplings of DSR and
gravity). The conditions of our calculation (that the deformation
is purely spatial) are then satisfied. It is possible that other
frameworks for DSR and its coupling to gravity satisfy this
condition.

Even within this framework other $U$ are possible.
For example with \be
U\circ (E,p)=(E,p(1+(\lambda p)^2))\; \ee one gets the
non-linear representation of Lorentz transformations:\bea
E'&=&\gamma[E-vp_x(1+(\lambda p)^2)]\nonumber\\
p_x'(1+(\lambda p')^2)&=&\gamma [p_x (1+(\lambda p)^2) -v
E]\nonumber\\
p_y'(1+(\lambda p')^2)&=&p_y (1+(\lambda p)^2) \nonumber\\
p_z'(1+(\lambda p')^2)&=&p_z (1+(\lambda p)^2)\; . \eea If we
restrict ourselves to transformations along the direction of
motion, in the limit $\lambda p\gg 1$ these may be written out
explicitly as: \bea E'&=&\gamma (E-vp (\lambda p)^2)\nonumber\\
p'&=&[\gamma (p^3 -vE\lambda^{-2})]^{1/3}\eea The HOD field theory
associated with it has Klein-Gordon equation: \be
[\partial_0^2-\partial_i^2(1+(\lambda\partial_i)^2)^2+m^2]\phi=0
\ee which is still within the requirements of our calculation.

But not all DSRs will do, even if they incorporate (\ref{theddr}).
For example we could have
taken \be E^2-k^2(1+(\lambda E)^2)^{\frac{2}{3}}=m^2 \ee which in
the massless limit is equivalent to (\ref{theddr}). It's realized by the
non-linear representation: \bea
E'&=&\gamma(E-vk_x(1+(\lambda E)^2)^{\frac{1}{3}})\\
k_x'&=&\frac{\gamma(k_x(1+(\lambda E)^2)^{\frac{1}{3}}-v E)}
{(1+\gamma^2\lambda^2(E-vk_x(1+(\lambda E)^2)^{\frac{1}{3}})^2)^{\frac{1}{3}}}\\
k_y'&=&\frac{k_y}
{(1+\gamma^2\lambda^2(E-vk_x(1+(\lambda E)^2)^{\frac{1}{3}})^2)^{\frac{1}{3}}}\\
k_z'&=&\frac{k_z} {(1+\gamma^2\lambda^2(E-vk_x(1+(\lambda
E)^2)^{\frac{1}{3}})^2)^{\frac{1}{3}}} \eea and the modified
Klein-Gordon theory: \be
[-\partial_0^2+\partial_i^2(1-\lambda\partial_0^2)^{\frac{1}{3}}+m^2]\phi=0\;
. \ee This is no longer consistent with the assumptions of the
calculation, i.e. complete decoupling between deformation and time
derivatives.

None of these DSR theories is ``doubly special'' in the sense that
it has an energy or momentum scale which is invariant under
Lorentz transformations. This could be easily implemented by
choosing DDRs of the form: \be \frac{E^2-p^2(1+(\lambda
p)^2)^2}{1-(L_{Pl}E)^2}=m^2\;. \ee Although the DDRs are the
correct ones the ensuing HOD field theory doesn't comply with the
assumptions of the calculation. It is possible, however, that a
modified calculation could be carried out and lead to
scale-invariance even for these theories.

\section{Conclusions}
If the speed of light, seen as a function of the wavelength, has a pole
of degree 2 at the origin, then the simplest adaptation of the
varying speed of sound mechanism for structure formation leads
to scale-invariant fluctuations. Additional minimal technical
assumptions on DSR and its coupling to gravity have to be made
(realized for example using~\cite{hod,rainbow}).
Other interpretations of DSR may or may not comply with the assumptions
of the calculation, and we encourage their proponents to carry out this
work. Here, for definiteness, we embedded the required DDRs into non-linear
representations of the Lorentz group~\cite{ljprd}, HOD field
theories~\cite{hod} and the rainbow metric~\cite{rainbow}, but it
may well be that this is not strictly necessary.

What practical advantages could this mechanism have  over inflation?
Foremost there's no need for reheating. The DSR behaviour studied in this
paper concerns regular matter, with the high energies experienced
by the early Universe triggering
VSL behavior. As the universe expands and cools this unusual
behavior disappears, leaving the Universed filled with standard
radiation engaged in ``business as usual'',
without the need for a ``decay'' or reheating. In other words there's
no esoteric matter here, merely regular matter behaving in an esoteric way.

Beyond this obvious practical advantage, we believe that the main
novelty of the scenario proposed here is that it connects better
with theories of phenomenology of quantum gravity, such
as the DSR arena.

I'd like to thank C. Armendariz-Picon, N. Barnaby, Y. Piao and the
participants of the workshop on non-commutative deformations of
relativity (ICMS, Edinburgh) for helpful comments.

\end{document}